\documentclass[pdflatex,sn-mathphys-num]{sn-jnl}

\usepackage{amsmath}
\usepackage{amssymb}
\usepackage{bm}

\usepackage{graphicx}
\usepackage{booktabs}
\usepackage{multirow}
\usepackage{diagbox}
\usepackage{xcolor}
\usepackage{colortbl}

\usepackage{tikz}
\usetikzlibrary{
    arrows.meta,
    calc,
    matrix,
    decorations.pathreplacing,
    positioning
}

\usepackage[braket,qm]{qcircuit}

\usepackage{subcaption}

\usepackage{algorithm}
\usepackage{algpseudocode}

\usepackage[utf8]{inputenc}
\usepackage[english]{babel}
\usepackage{hyperref}

\begin{document}

\title[Hybrid Quantum SchNet]{Variational Quantum Circuit Parameterization of SchNet: A Simulator-Based Feasibility Study for Conservative Molecular Force Fields}


\author[1]{\fnm{Hoang Anh} \sur{Nguyen}}

\author[2]{\fnm{Nhu Duc} \sur{Dinh}}
\equalcont{These authors contributed equally to this work.}

\author[3]{\fnm{Viet Hung} \sur{Tran}}
\equalcont{These authors contributed equally to this work.}

\author[1]{\fnm{Tu Uyen Le} \sur{Tu}}

\author[1,4]{\fnm{Tien Lam} \sur{Pham}}

\author*[1,4]{\fnm{Van Duy} \sur{Nguyen}}\email{duy.nguyenvan@phenikaa-uni.edu.vn}

\affil[1]{\orgdiv{Phenikaa School of Computing}, \orgname{Phenikaa University}, \city{Hanoi}, \postcode{12114}, \country{Vietnam}}
\affil[2]{\orgname{UFR PhITEM Universit´e Grenoble Alpes}, \orgaddress{\street{126 Rue de la Piscine}, \city{38400 Saint-Martin-d'Hères}, \country{France}}}

 \affil[3]{\orgdiv{Faculty of Materials Science}, \orgname{Phenikaa School of Engineering, Phenikaa University}, \city{Hanoi}, \postcode{12114}, \country{Vietnam}}

\affil[4]{\orgdiv{Phenikaa Institute for Advanced Study}, \orgname{Phenikaa University}, \city{Hanoi}, \postcode{12114}, \country{Vietnam}}

%
\date{\today}

\abstract{Machine-learning force fields provide a promising route for accelerating molecular simulation by replacing expensive quantum-chemical calculations with differentiable models of molecular energies and atomic forces. However, learning accurate and energy-conserving forces remains challenging, especially when the model must capture both global energy trends and local potential-energy gradients from limited data. In this work, we propose a Hybrid Quantum SchNet architecture that integrates variational quantum circuit modules into the continuous-filter SchNet framework. Quantum modules are inserted into the filter generator, atom-wise update, and readout transformations, allowing quantum-enhanced feature mappings to contribute to distance-dependent interactions and atomic energy prediction while preserving the energy-gradient formulation of forces. The model is evaluated on eight MD17 molecular systems using 1000 training configurations per molecule. Compared with energy-only training, joint energy--force supervision substantially improves both energy and force prediction accuracy. Compared with energy-only training, joint energy--force supervision substantially improves both energy and force prediction accuracy. Averaged over the benchmark, the energy MAE decreases from 2.567 to 0.593 kcal mol$^{-1}$, while the force MAE decreases from 16.340 to 1.540 kcal mol$^{-1}$ \AA$^{-1}$. Ablation experiments on ethanol further show that the performance of the hybrid model depends on the balance between quantum circuit width, circuit depth, and optimization stability. These results demonstrate that variational quantum circuits can be incorporated into neural force-field architectures and trained end-to-end to improve molecular energy and force prediction.}


\maketitle

\section{\label{sec:level1}Introduction}

\begin{figure}[t]
    \centering
    \includegraphics[width=0.95\linewidth]{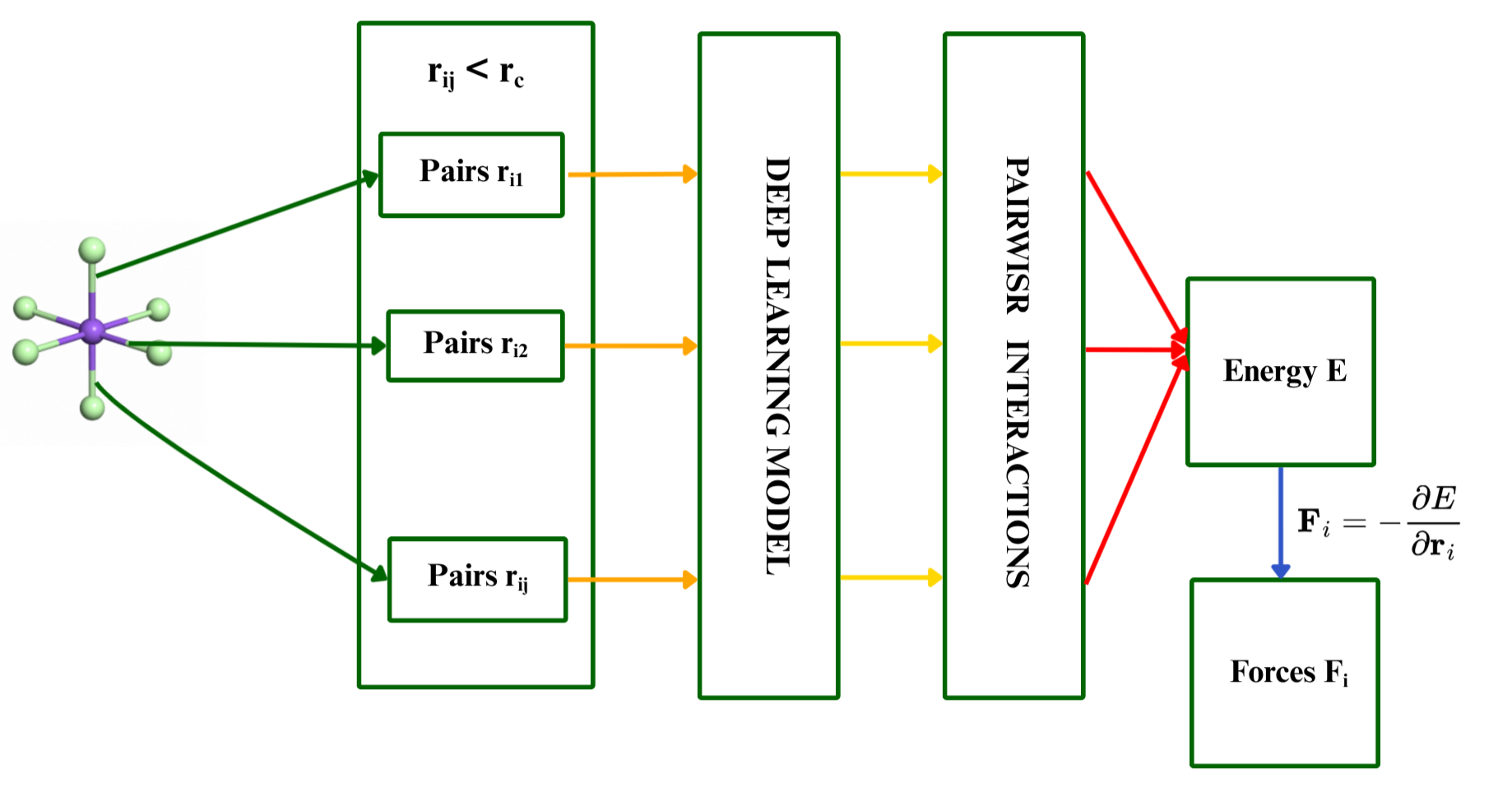}
    \caption{General formulation of the molecular energy and force prediction problem. An atomistic structure, represented by atomic numbers $\{Z_i\}$ and atomic coordinates $\{\mathbf{r}_i\}$, is mapped to a molecular energy $E$. Atomic forces are obtained as the negative gradient of the predicted energy with respect to atomic coordinates, $\mathbf{F}_i = -\nabla_{\mathbf{r}_i} E$, yielding an energy-conserving force field.
    }
\label{fig:problem_formulation}
\end{figure}

Accurate prediction of molecular energies and atomic forces is a central challenge in computational chemistry, molecular dynamics, and atomistic materials modeling. The potential energy surface determines the thermodynamic stability of molecular configurations, whereas the associated force field governs the time evolution of atomic coordinates. Conventional quantum-chemical methods, including density functional theory (DFT)~\cite{hohenberg_1964,kohn_1965}, provide high-fidelity reference data, but their computational cost restricts their direct application to long molecular-dynamics trajectories, large molecular systems, and broad configurational sampling. Machine-learning force fields address this bottleneck by learning differentiable surrogate models of the potential energy surface directly from atomistic data~\cite{unke_2021,behler_2007}. In this setting, physical consistency is not merely a numerical preference: atomic forces should be obtained as the negative gradient of the predicted energy with respect to atomic coordinates, yielding an energy-conserving force field suitable for molecular simulation~\cite{Chmiela_2017}. The energy--force prediction problem considered in this work is illustrated in Figure~\ref{fig:problem_formulation}.

The success of neural force fields depends strongly on how molecular structure is encoded into the model. A physically meaningful architecture should respect invariance of the total energy under global translations, rotations, and permutations of atoms of the same species, while producing forces that remain consistent with the underlying geometry. SchNet~\cite{SchNet_2018} is a representative architecture in this direction. It learns atom-wise latent representations through continuous-filter convolutions, where the filters are generated from interatomic distances. This design provides a natural mechanism for modeling local interactions while preserving the basic symmetries required for scalar molecular-energy prediction. Subsequent architectures, including PhysNet~\cite{PhysNet_2019}, DimeNet~\cite{DimeNet_2020}, PaiNN~\cite{PaiNN_2021}, NequIP~\cite{NequIP_2022}, and MACE~\cite{MACE_2022}, further demonstrate that chemically and geometrically informed inductive biases are essential for data-efficient learning of energies and forces.

Despite these advances, the trainable transformations used in most neural force fields remain entirely classical. Hybrid quantum-classical learning offers an alternative parameterization of nonlinear functions by embedding variational quantum circuits (VQCs) into neural architectures~\cite{Benedetti_2019,Cerezo_2021}. A VQC maps classical inputs to quantum states, evolves them through parameterized gates, and returns measured expectation values that are processed by classical layers. This paradigm is attractive because quantum circuits provide a distinct function class whose expressive behavior is controlled by circuit width, depth, entanglement structure, and measurement strategy. However, directly inserting quantum circuits into molecular force-field models is non-trivial. A useful hybrid model must preserve the physical constraints that make neural potentials reliable, including distance-based locality, symmetry consistency, differentiability with respect to atomic coordinates, and force prediction through energy gradients.

Hybrid quantum-classical models have recently been investigated in molecular and materials learning. Xia et al.~\cite{Xia_2020} studied hybrid quantum neural networks for electronic-structure calculations and potential-energy curves of small molecules. Quantum graph-based models have also been explored for molecular representation learning and materials property prediction~\cite{Vitz_2024,Piperno_2025}. More recently, Willow et al.~\cite{Willow_2025} proposed a hybrid quantum-classical machine-learning potential in which VQC-based modules are used within message-passing readout components for liquid silicon. These studies indicate that quantum modules can be incorporated into atomistic learning pipelines. Nevertheless, how VQCs should be embedded into the physically constrained components of a SchNet-like force field remains comparatively underexplored, especially when the target is simultaneous energy and force prediction under an energy-conserving formulation.

In this work, we propose Hybrid Quantum SchNet, a SchNet-compatible hybrid quantum-classical architecture for molecular energy and force prediction. The model retains the core energy-conserving structure of SchNet while introducing VQC modules into three trainable components: the continuous-filter generator, the atom-wise transformation in the interaction block, and the atom-wise readout network. The filter generator remains driven by radial distance features, the interaction block continues to aggregate local atomic environments, and the total molecular energy is computed by summing atom-wise energy contributions. Atomic forces are then obtained by differentiating the predicted energy with respect to atomic coordinates. In this way, the quantum modules modify the learnable nonlinear transformations without discarding the geometric and physical constraints of the underlying neural force field.

The contribution of this work is therefore methodological. We do not aim to outperform the strongest equivariant neural potentials, which are highly optimized classical architectures, but to establish a controlled framework for evaluating VQCs inside a physically constrained molecular force field. We evaluate the proposed model on the MD17 benchmark under energy-only and joint energy--force supervision. We further examine how quantum circuit width and depth affect ethanol energy prediction, thereby probing the trade-off between circuit capacity and trainability. The results show that joint force supervision is critical for accurate local gradients and that the proposed hybrid architecture can be optimized end-to-end on molecular energy and force data. These findings provide evidence that VQC modules can be studied meaningfully inside established neural force-field designs, rather than only as isolated readout layers or generic molecular predictors.

The remainder of this paper is organized as follows. Section~\ref{sec:background} introduces SchNet neural potentials and variational quantum circuits. Section~\ref{sec:methodology} presents the proposed Hybrid Quantum SchNet architecture. Section~\ref{sec:experiments} describes the dataset, model configuration, and training protocol. Section~\ref{sec:experiments_results} reports and discusses the numerical results, and Section~\ref{sec:conclusions} concludes the paper.

\section{\label{sec:background}Background}

\subsection{\label{sec:SchNet-neural-potentials}SchNet neural potentials}

SchNet is a continuous-filter convolutional neural network designed for atomistic systems~\cite{SchNet_2018}. It represents a molecule by its nuclear charges and atomic positions, $\mathcal{M}=\{(Z_i,\mathbf{r}_i)\}_{i=1}^{n},$
where $Z_i$ is the atomic number of atom $i$, $\mathbf{r}_i \in \mathbb{R}^{3}$ is its Cartesian coordinate, and $n$ is the number of atoms. The objective of a neural potential is to learn a scalar energy function $\hat{E}=\hat{E}(\mathbf{Z},\mathbf{R}),$
where $\mathbf{Z}=(Z_1,\ldots,Z_n)$ and $\mathbf{R}=(\mathbf{r}_1,\ldots,\mathbf{r}_n)$. Atomic forces are then obtained as the negative gradients of the predicted energy with respect to atomic positions, $\hat{\mathbf{F}}_i=-\partial \hat{E}/\partial \mathbf{r}_i$, for $i=1,\ldots,n$. This energy-based formulation produces conservative force fields and is therefore suitable for molecular dynamics applications.

In SchNet, each atom is first mapped to a trainable atom-type embedding,
\begin{equation}
    \mathbf{x}_i^{0}
    =
    \mathbf{a}_{Z_i},
\end{equation}
where $\mathbf{a}_{Z_i} \in \mathbb{R}^{F}$ is the learnable embedding vector associated with atomic number $Z_i$, and $F$ is the hidden feature dimension. At this stage, the representation encodes only the chemical identity of the corresponding atomic numbers. Geometric and environmental information is then introduced through a sequence of interaction blocks.

The main operation in SchNet is continuous-filter convolution. Unlike standard convolution on regular grids, molecular systems consist of atoms located at arbitrary positions. Accordingly, the convolutional filter is generated as a continuous function of interatomic geometry. At interaction layer $l$, the message aggregated by atom $i$ is written as
\begin{equation}
    \mathbf{c}_i^{l}
    =
    \sum_{j \in \mathcal{N}(i)}
    \tilde{\mathbf{x}}_j^{l}
    \odot
    W^{l}(d_{ij}),
\end{equation}
where $\mathcal{N}(i)$ is the neighbor set of atom $i$, $\odot$ denotes element-wise multiplication, $\tilde{\mathbf{x}}_j^{l}$ is an atom-wise transformed representation of atom $j$, and
\begin{equation}
    d_{ij}
    =
    \left\|
    \mathbf{r}_i
    -
    \mathbf{r}_j
    \right\|_2
\end{equation}
is the scalar interatomic distance. The filter $W^{l}(d_{ij})$ is generated from the distance between atoms and has the same feature dimension as the atom-wise representation.

In practice, the interatomic distance is expanded using radial basis functions before being passed to the filter-generating network:
\begin{equation}
    e_k(d_{ij})
    =
    \exp
    \left[
    -\gamma
    \left(
    d_{ij} - \mu_k
    \right)^2
    \right],
\end{equation}
where $\mu_k$ is the center of the $k$-th Gaussian basis function and $\gamma$ controls its width. The expanded radial feature vector is then mapped to a continuous filter,
\begin{equation}
    W^{l}(d_{ij})
    =
    f_{\theta}^{l}
    \left(
    \mathrm{RBF}(d_{ij})
    \right).
\end{equation}

After continuous-filter convolution, the representation of each atom is updated through a residual refinement,
\begin{equation}
    \mathbf{x}_i^{l+1}
    =
    \mathbf{x}_i^{l}
    +
    \mathbf{v}_i^{l},
\end{equation}
where $\mathbf{v}_i^{l}$ is obtained from the aggregated interaction feature $\mathbf{c}_i^{l}$. Stacking multiple interaction blocks allows the model to progressively incorporate local chemical environments and many-body effects into the atom-wise representations.

After the final interaction block, the molecular energy is computed from atom-wise energy contributions:
\begin{equation}
    \hat{E}_i
    =
    g_{\theta}
    \left(
    \mathbf{x}_i^{L}
    \right), 
    \qquad
    \hat{E}
    =
    \sum_{i=1}^{n}
    \hat{E}_i.
\end{equation}
The use of shared atom-wise operations and sum pooling ensures permutation invariance with respect to atom indexing. Since the filter depends on scalar interatomic distances rather than absolute coordinates or direction vectors, the predicted energy is invariant to global translations and rotations of the molecule. Consequently, the forces obtained by differentiating the energy are rotationally equivariant.

These properties make SchNet a physically structured neural potential: it learns atom-wise representations directly from atomic identities and interatomic distances while preserving the key symmetries required for molecular energy and force prediction.

\subsection{\label{sec:variational-quantum-circuits}Variational quantum circuits}

Variational quantum circuits (VQCs) are parameterized quantum models that can be embedded into hybrid quantum-classical learning architectures~\cite{Benedetti_2019,Cerezo_2021}. A VQC typically consists of three stages: classical data encoding into a quantum state, parameterized quantum evolution, and measurement of quantum observables. Given a $d$-dimension classical input vector $\mathbf{h}=(h_1,h_2,\ldots,h_d)\in\mathbb{R}^{d},$ the first step is to encode the input into a $q$-qubit quantum state. Since the classical input dimension may differ from the number of qubits, a classical projection layer can be used to map the input vector to a $q$-dimensional quantum input:
\begin{equation}
    \mathbf{z}
    =
    \phi
    \left(
    W_{\mathrm{in}}\mathbf{h}
    +
    \mathbf{b}_{\mathrm{in}}
    \right),
    \qquad
    \mathbf{z}
    \in
    \mathbb{R}^{q},
\end{equation}
where $W_{\mathrm{in}}$ and $\mathbf{b}_{\mathrm{in}}$ are trainable classical parameters and $\phi(\cdot)$ is an activation function.

In angle encoding, the components of $\mathbf{z}$ are used as rotation angles. Using $\mathrm{R}_{Y}$ rotations, the encoded state can be written as
\begin{equation}
    \ket{\psi_{\mathrm{enc}}(\mathbf{z})}
    =
    U_{\mathrm{enc}}(\mathbf{z})
    \ket{0}^{\otimes q},
\end{equation}
with
\begin{equation}
    U_{\mathrm{enc}}(\mathbf{z})
    =
    \prod_{k=1}^{q}
    \mathrm{R}_{Y}(z_k).
\end{equation}

After data encoding, the quantum state is processed by a trainable ansatz,
\begin{equation}
    \ket{\psi(\mathbf{z};\boldsymbol{\omega})}
    =
    U_{\mathrm{ans}}(\boldsymbol{\omega})
    U_{\mathrm{enc}}(\mathbf{z})
    \ket{0}^{\otimes q},
\end{equation}
where $\boldsymbol{\omega}$ denotes the trainable quantum parameters. A common ansatz is composed of repeated layers of parameterized single-qubit rotations and entangling gates. In this work, the VQC modules use strongly entangling layers~\cite{schuld_2020}, where each layer applies trainable single-qubit rotation gates followed by two-qubit entangling operations. The single-qubit rotation can be written as
\begin{equation}
    \mathrm{Rot}(\alpha,\beta,\gamma)
    =
    \mathrm{R}_{Z}(\alpha)\mathrm{R}_{Y}(\beta)\mathrm{R}_{Z}(\gamma).
\end{equation}

The circuit output is obtained by measuring expectation values of selected observables. Using Pauli-$Z$ measurements over all qubits gives the quantum feature vector
\begin{equation}
    \mathbf{m}
    =
    \left[
    \langle Z_1 \rangle,
    \langle Z_2 \rangle,
    \ldots,
    \langle Z_q \rangle
    \right]^{\top},
\end{equation}
where
\begin{equation}
    \langle Z_k \rangle
    =
    \bra{\psi(\mathbf{z};\boldsymbol{\omega})}
    Z_k
    \ket{\psi(\mathbf{z};\boldsymbol{\omega})}.
\end{equation}
The measured expectation values are classical quantities and can therefore be passed to subsequent classical layers.

A hybrid VQC module can therefore be viewed as a trainable nonlinear map from a classical input space to a classical output space, $\mathrm{VQC}_{\Omega}: \mathbb{R}^{d_{\mathrm{in}}} \rightarrow \mathbb{R}^{d_{\mathrm{out}}}$,
where $\Omega$ contains both classical projection parameters and quantum circuit parameters. After the measured quantum feature vector is obtained, a classical output projection maps it to the required output dimension:
\begin{equation}
    \mathbf{y}
    =
    W_{\mathrm{out}}\mathbf{m}
    +
    \mathbf{b}_{\mathrm{out}},
    \qquad
    \mathbf{y}
    \in
    \mathbb{R}^{d_{\mathrm{out}}}.
\end{equation}
Here, $W_{\mathrm{out}}$ and $\mathbf{b}_{\mathrm{out}}$ are trainable classical parameters.

In hybrid quantum-classical neural networks, VQC modules are not limited to standalone predictors. They can also be inserted into larger neural architectures as trainable nonlinear components. This makes them suitable for parameterizing selected mappings inside SchNet, such as the filter-generating network, atom-wise refinement layers, and readout function, while preserving the overall energy-conserving structure of the neural potential.


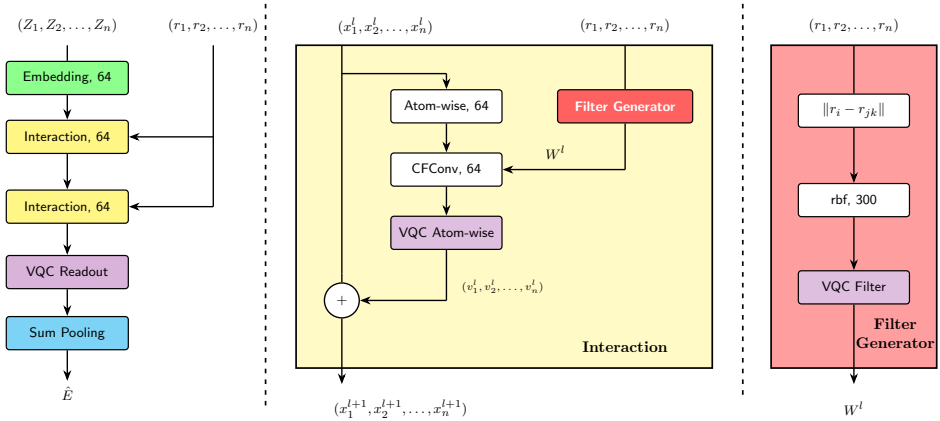
\begin{figure}[!htbp]
    \centering
    \resizebox{0.95\linewidth}{!}{%
    \begin{tikzpicture}[
        font=\small\sffamily,
        >=Stealth,
        block/.style={
            draw=black, rounded corners=2pt,
            minimum width=2.8cm, minimum height=0.76cm,
            align=center, line width=0.8pt
        },
        sblock/.style={
            draw=black, rounded corners=2pt,
            minimum width=2.55cm, minimum height=0.76cm,
            align=center, line width=0.8pt
        },
        arr/.style={->, line width=0.9pt},
        lin/.style={line width=0.9pt},
        sep/.style={dashed, line width=1.1pt, dash pattern=on 3pt off 4pt}
    ]

    \node at (0,6.50)    {$(Z_1,Z_2,\ldots,Z_n)$};
    \node at (3.35,6.50) {$(r_1,r_2,\ldots,r_n)$};

    \node[block, fill=green!45]  (emb)  at (0,5.30) {Embedding, 64};
    \node[block, fill=yellow!60] (int1) at (0,3.95) {Interaction, 64};
    \node[block, fill=yellow!60] (int2) at (0,2.35) {Interaction, 64};
    \node[block, fill=violet!30] (vqcr) at (0,0.85) {VQC Readout};
    \node[block, fill=cyan!45]   (pool) at (0,-0.55){Sum Pooling};

    \draw[lin] (0,6.05) -- (emb.north);
    \draw[arr] (emb)  -- (int1);
    \draw[arr] (int1) -- (int2);
    \draw[arr] (int2) -- (vqcr);
    \draw[arr] (vqcr) -- (pool);
    \draw[arr] (pool.south) -- ++(0,-0.70) node[below]{$\hat{E}$};

    \draw[lin] (3.35,6.05) -- (3.35,2.35);
    \draw[arr] (3.35,3.95) -- (int1.east);
    \draw[arr] (3.35,2.35) -- (int2.east);

    \draw[sep] (4.55,7.00) -- (4.55,-2.28);

    \draw[draw=black, fill=yellow!28, line width=1.1pt]
        (5.25,-1.35) rectangle (14.80,6.05);

    \node at (7.30,6.50)  {$(x_1^l,x_2^l,\ldots,x_n^l)$};
    \node at (12.80,6.50) {$(r_1,r_2,\ldots,r_n)$};

    \node[sblock, fill=white]     (atom) at (8.70,4.65) {Atom-wise, 64};
    \node[sblock, fill=white]     (cf)   at (8.70,3.20) {CFConv, 64};
    \node[sblock, fill=violet!25] (vqca) at (8.70,1.75) {VQC Atom-wise};

    \node[draw=black, rounded corners=2pt, fill=red!62, text=white,
          minimum width=3.1cm, minimum height=0.76cm,
          align=center, line width=0.8pt]
        (fgen) at (12.80,4.65) {\textbf{Filter Generator}};

    \node[circle, draw=black, fill=white, minimum size=8mm,
          inner sep=0pt, line width=0.9pt] (plus) at (6.30,0.20) {$+$};

    \draw[lin] (6.30,6.05) -- (6.30,0.82);
    \draw[lin] (6.30,0.82) -- (plus.north);
    \draw[lin] (6.30,5.40) -- (8.70,5.40);
    \draw[arr] (8.70,5.40) -- (atom.north);
    \draw[arr] (atom.south) -- (cf.north);
    \draw[arr] (cf.south)   -- (vqca.north);

    \draw[lin] (vqca.south) -- (8.70,0.20);
    \draw[arr] (8.70,0.20)  -- (plus.east);

    \node[font=\footnotesize\itshape, anchor=west] at (8.85,0.52)
        {\;$(v_1^l,v_2^l,\ldots,v_n^l)$};

    \draw[lin] (12.80,6.05) -- (fgen.north);
    \draw[lin] (fgen.south) -- (12.80,3.20);
    \draw[arr] (12.80,3.20) -- (cf.east);
    \node[font=\small\itshape] at (11.20,3.60) {$W^l$};

    \draw[arr] (plus.south) -- (6.30,-1.72);
    \node at (7.65,-2.28) {$(x_1^{l+1},x_2^{l+1},\ldots,x_n^{l+1})$};
    \node[font=\bfseries\normalsize] at (12.80,-0.85) {Interaction};

    \draw[sep] (15.50,7.00) -- (15.50,-2.10);

    \draw[draw=black, fill=red!38, line width=1.1pt]
        (16.15,-1.35) rectangle (19.95,6.05);

    \node at (18.05,6.50) {$(r_1,r_2,\ldots,r_n)$};

    \node[sblock, fill=white]     (dist) at (18.05,4.55) {$\|r_i - r_{jk}\|$};
    \node[sblock, fill=white]     (rbf)  at (18.05,2.50) {rbf, 300};
    \node[sblock, fill=violet!25] (vqcf) at (18.05,0.5) {VQC Filter};

    \draw[lin] (18.05,6.05) -- (dist.north);
    \draw[arr] (dist.south) -- (rbf.north);
    \draw[arr] (rbf.south)  -- (vqcf.north);
    \draw[arr] (vqcf.south) -- (18.05,-1.72);

    \node[font=\bfseries\normalsize, align=center] at (19.0,-0.55) {Filter\\Generator};

    \node at (18.05,-2.28) {$W^l$};


    \end{tikzpicture}
    }
    \caption{
    Overview of the proposed Hybrid Quantum SchNet architecture.
    Atomic numbers are mapped to atom-wise embeddings, while atomic positions
    are used to compute interatomic distances and radial basis features.
    The atom-wise representations are refined through interaction blocks using
    quantum-generated continuous filters and hybrid VQC atom-wise updates.
    A hybrid VQC readout maps the final atom-wise representations to atomic
    energy contributions, which are summed to obtain the predicted molecular
    energy $\hat{E}$.
    Atomic forces are computed as negative gradients of $\hat{E}$ with respect
    to atomic positions.
    }
    \label{fig:SchNet_architecture}
\end{figure}

\section{\label{sec:methodology}Proposed Hybrid Quantum SchNet}
\subsection{\label{sec:model-overview}Model overview}

The proposed Hybrid Quantum SchNet follows the energy-conserving formulation of SchNet while introducing hybrid variational quantum circuit modules into selected trainable transformations. Given a molecular configuration $\mathcal{M}=\left\{(Z_i,\mathbf{r}_i)\right\}_{i=1}^{n},$ the model predicts a scalar molecular energy $\hat{E}=\hat{E}_{\Theta}(\mathbf{Z},\mathbf{R}),$
where $\Theta$ denotes the full set of trainable parameters. Atomic forces are obtained from the predicted energy as $\hat{\mathbf{F}}_i=-\nabla_{\mathbf{r}_i}\hat{E}_{\Theta}(\mathbf{Z},\mathbf{R})$, for $i=1,\ldots,n$.

Compared with the classical SchNet model, the proposed architecture modifies three trainable components: the distance-based filter generator, the post-convolution atom-wise refinement, and the final atom-wise readout. These components are parameterized by hybrid VQC modules, denoted as
$\mathrm{VQC}_{\mathrm{filter}}$,
$\mathrm{VQC}_{\mathrm{atomwise}}$, and
$\mathrm{VQC}_{\mathrm{readout}}$, respectively.
The overall architecture is shown in Fig.~\ref{fig:SchNet_architecture}.

The design choice is motivated by the different roles of these three transformations. The filter generator controls how geometric distance information is converted into pairwise interaction filters. The atom-wise refinement module processes the local environment information after continuous-filter convolution. The readout module converts the final atomic representation into an atomic energy contribution. Thus, the proposed model does not replace the physical structure of SchNet; rather, it reparameterizes selected nonlinear mappings inside that structure using hybrid quantum-classical transformations.

\subsection{\label{sec:generic-vqc-module}Hybrid variational quantum circuit module}
\begin{figure}[!htbp]
    \centering
    \includegraphics[width=0.98\linewidth]{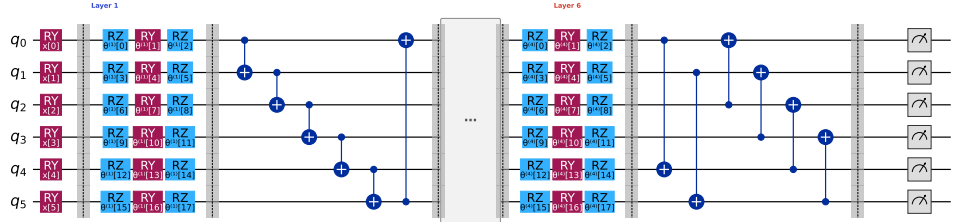}
    \caption{
    Variational quantum circuit used in the hybrid quantum modules.
    Classical input features are projected to a six-dimensional quantum input vector and encoded into six qubits using $\mathrm{R}_{Y}$ angle encoding.
    The encoded state is processed by strongly entangling layers composed of parameterized single-qubit $\mathrm{Rot}$ gates, where $\mathrm{Rot}(\alpha,\beta,\gamma)=\mathrm{R}_{Z}(\alpha)\mathrm{R}_{Y}(\beta)\mathrm{R}_{Z}(\gamma)$, followed by two-qubit entangling gates.
    The quantum output is obtained from Pauli-$Z$ expectation-value measurements over all qubits and mapped to the required output dimension by a classical linear projection.
    }
    \label{fig:vqc-circuit}
\end{figure}

The hybrid quantum modules in the proposed architecture use the generic VQC formulation introduced in Section~\ref{sec:variational-quantum-circuits}. In this work, each VQC instance is implemented with six qubits, $\mathrm{R}_{Y}$ angle encoding, strongly entangling layers, and Pauli-$Z$ expectation-value measurements, as illustrated in Fig.~\ref{fig:vqc-circuit}.

For a classical input feature vector $\mathbf{h}\in\mathbb{R}^{d_{\mathrm{in}}}$, a classical projection layer first maps the input to a six-dimensional quantum input vector:
\begin{equation}
    \tilde{\mathbf{z}}
    =
    W_{\mathrm{in}}\mathbf{h}
    +
    \mathbf{b}_{\mathrm{in}},
    \qquad
    \mathbf{z}
    =
    \pi \tanh(\tilde{\mathbf{z}}),
    \qquad
    \mathbf{z}\in\mathbb{R}^{6}.
\end{equation}
The $\tanh$ nonlinearity bounds the projected values to $[-1,1]$, and the factor $\pi$ rescales them to $[-\pi,\pi]$ before angle encoding. After quantum processing and Pauli-$Z$ measurements, a classical output projection maps the measured quantum features to the required output dimension. This input--output projection allows a common circuit template to operate on feature vectors with different dimensions.

Although the three quantum modules share the same circuit structure, they are used for different transformations in the SchNet pipeline:
\begin{equation}
\begin{aligned}
    \mathrm{VQC}_{\mathrm{filter}}&: \mathbb{R}^{K} \rightarrow \mathbb{R}^{F},\\
    \mathrm{VQC}_{\mathrm{atomwise}}&: \mathbb{R}^{F} \rightarrow \mathbb{R}^{F},\\
    \mathrm{VQC}_{\mathrm{readout}}&: \mathbb{R}^{F} \rightarrow \mathbb{R}.
\end{aligned}
\end{equation}
Here, $K$ is the number of radial basis functions and $F$ is the hidden feature dimension. Each module has its own trainable parameters and therefore learns a task-specific transformation, even though the underlying circuit design is shared.

\subsection{\label{sec:atomic-embedding}Atomic embedding}

The atomic embedding stage is kept identical to SchNet. Each atomic number is mapped to a trainable feature vector,
\begin{equation}
    \mathbf{x}_i^{0}
    =
    \mathbf{a}_{Z_i},
    \qquad
    \mathbf{a}_{Z_i}\in\mathbb{R}^{F}.
\end{equation}
In this work, we set $F$ = 64. The embedding layer provides the initial atom-wise representation before geometry-dependent interaction blocks are applied.

This stage is intentionally left classical because it only assigns a learnable representation to each atom type. The quantum modules are introduced later, after distance information and local environment information become available. This separation keeps the input representation simple while allowing the quantum modules to act on physically meaningful continuous features.

\subsection{\label{sec:hybrid-filter-generator}Hybrid quantum filter generator}

For each pair $(i,j)$ with $j\in\mathcal{N}(i)$, the interatomic distance is computed from the atomic coordinates and expanded into radial basis features:
\begin{equation}
    d_{ij}
    =
    \|\mathbf{r}_i-\mathbf{r}_j\|_2,
    \qquad
    \mathbf{g}_{ij}
    =
    \mathrm{RBF}(d_{ij}).
\end{equation}
The Gaussian basis expansion follows the definition in Section~\ref{sec:SchNet-neural-potentials}.

The filter generator is the first quantum-enhanced component. Its role is to transform the radial distance feature $\mathbf{g}_{ij}$ into a pairwise continuous filter used in the convolution. In the proposed model, this mapping is parameterized by a hybrid VQC module:
\begin{equation}
    W_{ij}^{l}
    =
    \mathrm{VQC}_{\mathrm{filter}}^{l}
    \left(
    \mathbf{g}_{ij}
    \right),
    \qquad
    W_{ij}^{l}\in\mathbb{R}^{F}.
\end{equation}

This modification targets the geometry-to-interaction mapping of SchNet. Since $W_{ij}^{l}$ determines how the representation of atom $j$ contributes to the update of atom $i$, the expressiveness of the filter generator directly affects the learned interaction pattern. By using $\mathrm{VQC}_{\mathrm{filter}}$, the distance-dependent filter is produced by a hybrid quantum-classical nonlinear transformation, while the input remains based only on scalar distances. Consequently, the quantum replacement changes the parameterization of the filter function without altering the geometric invariances of the original distance-based construction.

\subsection{\label{sec:hybrid-interaction-block}Hybrid quantum interaction block}

At interaction layer $l$, each atom representation is first transformed by a shared atom-wise linear layer:
\begin{equation}
    \tilde{\mathbf{x}}_i^{l}
    =
    W_{\mathrm{aw}}^{l}
    \mathbf{x}_i^{l}
    +
    \mathbf{b}_{\mathrm{aw}}^{l}.
\end{equation}
The continuous-filter convolution is then computed using the quantum-generated filters:
\begin{equation}
    \mathbf{c}_i^{l}
    =
    \sum_{j\in\mathcal{N}(i)}
    \tilde{\mathbf{x}}_j^{l}
    \odot
    W_{ij}^{l}.
\end{equation}

The vector $\mathbf{c}_i^{l}$ is the aggregated local environment representation of atom $i$ at layer $l$. It combines the transformed neighboring atom features with the distance-dependent filters produced by $\mathrm{VQC}_{\mathrm{filter}}$. This vector therefore contains both chemical information from neighboring atom representations and geometric information from interatomic distances.

The second quantum-enhanced component is applied after this aggregation. The local interaction feature is passed to a hybrid VQC atom-wise module:
\begin{equation}
    \mathbf{v}_i^{l}
    =
    \mathrm{VQC}_{\mathrm{atomwise}}^{l}
    \left(
    \mathbf{c}_i^{l}
    \right),
    \qquad
    \mathbf{v}_i^{l}\in\mathbb{R}^{F}.
\end{equation}
The output $\mathbf{v}_i^{l}$ is used as the residual update:
\begin{equation}
    \mathbf{x}_i^{l+1}
    =
    \mathbf{x}_i^{l}
    +
    \mathbf{v}_i^{l}.
\end{equation}

The role of $\mathrm{VQC}_{\mathrm{atomwise}}$ is different from that of $\mathrm{VQC}_{\mathrm{filter}}$. While the filter module acts on pairwise distance features, the atom-wise module acts on the aggregated local environment feature of each atom. This allows the quantum module to refine atom representations after neighbor information has been collected. Since the same atom-wise module is shared across atoms, the interaction block remains permutation equivariant.

\subsection{\label{sec:hybrid-readout}Hybrid quantum readout}

After $L$ interaction blocks, each atom has a final representation $\mathbf{x}_i^{L}$ that encodes its chemical identity and local environment. The third quantum-enhanced component is the readout network, which maps this final atom-wise representation to an atomic energy contribution:
\begin{equation}
    \hat{E}_i
    =
    \mathrm{VQC}_{\mathrm{readout}}
    \left(
    \mathbf{x}_i^{L}
    \right),
    \qquad
    \hat{E}_i\in\mathbb{R}.
\end{equation}
The total molecular energy is obtained by summing over atoms:
\begin{equation}
    \hat{E}
    =
    \sum_{i=1}^{n}
    \hat{E}_i.
\end{equation}

The readout module has a different role from the filter and atom-wise modules. It does not generate pairwise interaction filters or refine hidden representations. Instead, it maps the final atomic feature vector to a scalar energy contribution. Using a hybrid VQC at this stage tests whether the quantum module can serve as the final nonlinear energy mapping while preserving the extensive atom-wise energy decomposition of SchNet.

\subsection{\label{sec:energy-force-training}Energy-force training objective}

Forces are computed from the predicted energy as $\hat{\mathbf{F}}_i=-\nabla_{\mathbf{r}_i}\hat{E}$.
The model is trained using the combined energy--force loss
\begin{equation}
    \mathcal{L}
    =
    \lambda_E \left| \hat{E} - E \right|^2
    +
    \frac{\lambda_F}{3n}
    \sum_{i=1}^{n} \sum_{\alpha=1}^{3}
    \left|
    -\frac{\partial \hat{E}}{\partial r_{i,\alpha}}
    -
    F_{i,\alpha}
    \right|^2,
\end{equation}
where $n$ is the number of atoms and $\lambda_E$ and $\lambda_F$ control the relative weights of energy and force supervision.

The energy term supervises the scalar potential energy surface, while the force term supervises its local derivatives with respect to atomic coordinates. Including force labels provides additional geometric information about the local shape of the potential energy surface. Classical and quantum parameters are optimized jointly in the same computational graph.

\section{\label{sec:experiments}Experiments}

\subsection{\label{sec:dataset_protocols}Dataset and splitting protocol}

We evaluate the proposed Hybrid Quantum SchNet model on the MD17 molecular dynamics benchmark. 
The experiments are conducted on eight molecular systems: benzene, toluene, malonaldehyde, salicylic acid, aspirin, ethanol, uracil, and naphthalene. 
Each molecule is treated as an independent learning task. 
For each molecular configuration, the input consists of the atomic numbers $\{Z_i\}_{i=1}^{n}$ and Cartesian coordinates $\{\mathbf{r}_i\}_{i=1}^{n}$. 
The learning targets are the molecular energy $E$ and the atomic forces $\{\mathbf{F}_i\}_{i=1}^{n}$.

For the main MD17 experiments, we use $N=1000$ training configurations for each molecule. 
A separate validation set is used to monitor the training process and select the model checkpoint, while the remaining configurations are used for testing. 
The same data split is used when comparing the energy-only and energy+force training settings, so that the observed differences are due to the training objective rather than changes in the data partition. 
Energy errors are reported in kcal mol$^{-1}$, and force errors are reported in kcal mol$^{-1}$ \AA$^{-1}$.

The model is trained and evaluated separately for each molecule. 
This protocol follows the standard molecular force-field learning setting, where a separate potential energy surface is learned for each molecular system. 
No information is shared across different molecules during training.

\subsection{\label{sec:model_configuration}Model configuration}

All experiments use the full Hybrid Quantum SchNet configuration described in Section~\ref{sec:methodology}. 
In this configuration, hybrid variational quantum circuit modules are inserted into three trainable components of the SchNet architecture: the distance-based filter generator, the atom-wise interaction update, and the atom-wise readout network. 
The filter module maps radial distance features to continuous convolutional filters, the atom-wise module refines the aggregated local environment representation, and the readout module maps final atom-wise representations to atomic energy contributions.

The present study focuses on this full hybrid configuration as a representative SchNet-compatible quantum-classical neural potential. 
The results should therefore be interpreted as an initial feasibility and trainability evaluation of the proposed hybrid architecture, rather than as a claim that this particular placement of quantum modules is optimal. 
A detailed placement ablation, in which the filter generator, atom-wise update, and readout modules are replaced separately, is left for future work.

The atom-wise hidden feature dimension is fixed to $F=64$. 
The atomic embedding layer maps each atomic number to a trainable feature vector in $\mathbb{R}^{64}$. 
Interatomic distances are expanded using Gaussian radial basis functions before being passed to the filter generator. 
The same radial basis expansion and neighbor construction are used for all training settings to ensure a controlled comparison.

The molecular energy is predicted as a sum of atom-wise energy contributions,
\begin{equation}
    \hat{E}
    =
    \sum_{i=1}^{n}
    \hat{E}_i ,
\end{equation}
and atomic forces are obtained by differentiating the predicted energy with respect to atomic coordinates,
\begin{equation}
    \hat{\mathbf{F}}_i
    =
    -
    \nabla_{\mathbf{r}_i}
    \hat{E}.
\end{equation}
This construction preserves the energy-conserving force-field formulation throughout all experiments.

\subsection{\label{sec:quantum_circuit_configurations}Quantum circuit configurations}

Each hybrid VQC module follows the circuit structure introduced in Section~\ref{sec:generic-vqc-module}. 
A classical input feature vector is first projected to a $q$-dimensional quantum input vector,
\begin{equation}
    \tilde{\mathbf{z}}
    =
    W_{\mathrm{in}}\mathbf{h}
    +
    \mathbf{b}_{\mathrm{in}},
    \qquad
    \mathbf{z}
    =
    \pi \tanh(\tilde{\mathbf{z}}),
    \qquad
    \mathbf{z}\in\mathbb{R}^{q}.
\end{equation}
The bounded vector $\mathbf{z}$ is encoded into a $q$-qubit quantum state using $\mathrm{R}_{Y}$ angle encoding. 
The encoded state is then processed by strongly entangling layers composed of trainable single-qubit $\mathrm{Rot}$ gates and two-qubit entangling gates. 
The circuit output is obtained from Pauli-$Z$ expectation-value measurements on all qubits and is mapped to the required output dimension by a classical linear projection.

For the main MD17 results, we use the default full hybrid configuration with $q=6$ qubits. 
To examine the sensitivity of the model to quantum circuit size, we additionally vary the number of qubits and the number of strongly entangling layers in a controlled experiment. 
Specifically, the number of qubits is varied as $q \in \{3,4,5,6\},$ and the number of quantum layers is varied as  $L_q \in \{1,2,3,4,5,6\}$.
This qubit-depth analysis is performed on ethanol as a representative small molecule, and the resulting energy MAE is used to study how circuit size affects predictive performance.

The quantum circuits are evaluated using expectation values of Pauli-$Z$ observables. 
Unless otherwise stated, the reported experiments use exact expectation values from quantum circuit simulation and do not include finite-shot sampling noise. 
Thus, the reported results isolate the effect of the hybrid quantum-classical parameterization from the additional stochasticity that would arise on finite-shot quantum hardware.

\subsection{\label{sec:training_details}Training details}

The model is trained by minimizing the combined energy-force objective defined in Section~\ref{sec:energy-force-training}. 
For a molecule with $n$ atoms, the loss for one molecular configuration is written as
\begin{equation}
    \mathcal{L}
    =
    \lambda_E
    \left|
    \hat{E}-E
    \right|^2
    +
    \frac{\lambda_F}{3n}
    \sum_{i=1}^{n}
    \sum_{\alpha=1}^{3}
    \left|
    \hat{F}_{i,\alpha}
    -
    F_{i,\alpha}
    \right|^2 ,
\end{equation}
where $\lambda_E$ and $\lambda_F$ control the relative weights of the energy and force terms. 
The predicted forces are computed as
\begin{equation}
    \hat{F}_{i,\alpha}
    =
    -
    \frac{\partial \hat{E}}{\partial r_{i,\alpha}}.
\end{equation}

Two training settings are considered. 
The first is the energy-only setting, where the model is trained using only the molecular energy target,
\begin{equation}
    \lambda_E = 1.0,
    \qquad
    \lambda_F = 0.0.
\end{equation}
The second is the joint energy-force setting, where both energy and force labels are used,
\begin{equation}
    \lambda_E = 0.01,
    \qquad
    \lambda_F = 1.0.
\end{equation}
The energy-only setting evaluates whether the model can learn the global potential energy surface from scalar energy labels alone. 
The joint energy-force setting evaluates whether local gradient information improves the learned potential and the resulting force predictions.

All models are optimized using Adam. 
The batch size is fixed to 32. 
Because the classical neural-network parameters and quantum circuit angles have different numerical roles in the hybrid architecture, we use separate learning rates for the two parameter groups. 
The classical parameters are optimized with learning rate $10^{-3}$, while the quantum parameters are optimized with learning rate $10^{-4}$. 
An exponential learning-rate decay with decay factor $0.96$ is applied during training. 
An exponential moving average of the model parameters is used to stabilize evaluation after optimization steps.

Classical and quantum parameters are optimized jointly within the same computational graph. 
Since atomic forces are obtained by differentiating the predicted energy with respect to atomic coordinates, the training procedure requires gradients through both the SchNet computation and the hybrid VQC modules.

\subsection{\label{sec:evaluation_metrics}Evaluation metrics}

The predictive performance is evaluated using mean absolute error (MAE) and the coefficient of determination $R^2$. 
For a test set with $M$ molecular configurations, the energy MAE is defined as
\begin{equation}
    \mathrm{MAE}_{E}
    =
    \frac{1}{M}
    \sum_{s=1}^{M}
    \left|
    \hat{E}^{(s)}
    -
    E^{(s)}
    \right|.
\end{equation}

The force MAE is computed over all atoms and Cartesian components,
\begin{equation}
    \mathrm{MAE}_{F}
    =
    \frac{1}{M}
    \sum_{s=1}^{M}
    \frac{1}{3n_s}
    \sum_{i=1}^{n_s}
    \sum_{\alpha=1}^{3}
    \left|
    \hat{F}_{i,\alpha}^{(s)}
    -
    F_{i,\alpha}^{(s)}
    \right|,
\end{equation}
where $n_s$ is the number of atoms in configuration $s$. 
For each molecule in MD17, $n_s$ is constant across the trajectory.


\section{\label{sec:experiments_results}Results}

\begin{table*}[!t]
\centering
\caption{Comparison of MAEs for the reproduced classical SchNet baseline and Hybrid Quantum SchNet on MD17 using $N=1000$ training configurations. Energy and force MAEs are reported in kcal mol$^{-1}$ and kcal mol$^{-1}$ \AA$^{-1}$, respectively. Both models were trained and evaluated under the same experimental protocol.}
\label{tab:hybrid_SchNet_md17_results}
\resizebox{\textwidth}{!}{%
\begin{tabular}{llcccc}
\hline
\multirow{2}{*}{\textbf{Molecule}} & \multirow{2}{*}{\textbf{Target}}
& \multicolumn{2}{c}{\textbf{Energy-only training}}
& \multicolumn{2}{c}{\textbf{Energy+force training}} \\
& & \textbf{Reproduced SchNet} & \textbf{Hybrid Quantum SchNet}
  & \textbf{Reproduced SchNet} & \textbf{Hybrid Quantum SchNet} \\
\hline
\multirow{2}{*}{Benzene}
& Energy & 1.19 & 0.579 & 0.08 & 0.087 \\
& Forces & 14.12 & 6.104 & 0.31 & 0.439 \\
\hline
\multirow{2}{*}{Toluene}
& Energy & 2.95 & 2.884 & 0.12 & 0.642 \\
& Forces & 22.31 & 17.848 & 0.57 & 1.942 \\
\hline
\multirow{2}{*}{Malonaldehyde}
& Energy & 2.03 & 2.025 & 0.13 & 0.667 \\
& Forces & 20.41 & 16.169 & 0.66 & 1.675 \\
\hline
\multirow{2}{*}{Salicylic acid}
& Energy & 3.27 & 3.625 & 0.20 & 0.452 \\
& Forces & 23.21 & 20.007 & 0.85 & 1.373 \\
\hline
\multirow{2}{*}{Aspirin}
& Energy & 4.20 & 4.111 & 0.37 & 1.509 \\
& Forces & 23.54 & 19.840 & 1.35 & 2.707 \\
\hline
\multirow{2}{*}{Ethanol}
& Energy & 0.93 & 1.159 & 0.08 & 0.257 \\
& Forces & 6.56 & 8.755 & 0.39 & 0.763 \\
\hline
\multirow{2}{*}{Uracil}
& Energy & 2.26 & 2.381 & 0.14 & 0.421 \\
& Forces & 20.08 & 19.796 & 0.56 & 1.571 \\
\hline
\multirow{2}{*}{Naphthalene}
& Energy & 3.58 & 3.772 & 0.16 & 0.712 \\
& Forces & 25.36 & 22.200 & 0.58 & 1.854 \\
\hline
\end{tabular}%
}

\vspace{0.3em}
\begin{minipage}{0.98\textwidth}
\footnotesize
\end{minipage}
\end{table*}

\subsection{\label{sec:experimental_setup}Experimental setup}

\begin{figure}[!htbp]
    \centering
    \begin{subfigure}[t]{0.49\linewidth}
        \centering
        \includegraphics[width=\linewidth]{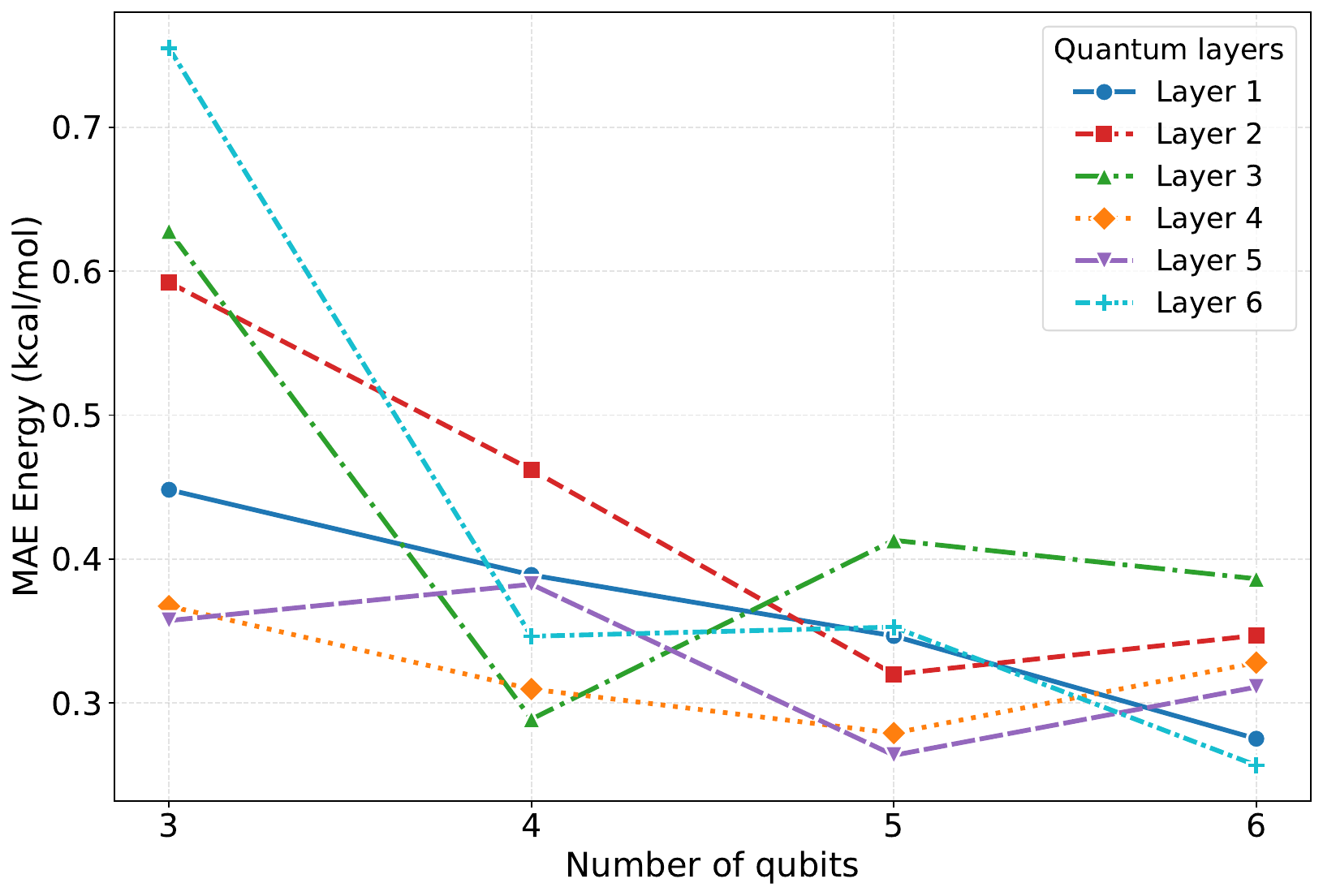}
        \caption{MAE energy experiments of ethanol across qubit and layer configurations.}
        \label{fig:ethanol_mae_qubits_layers}
    \end{subfigure}
    \hfill
    \begin{subfigure}[t]{0.49\linewidth}
        \centering
        \includegraphics[width=\linewidth]{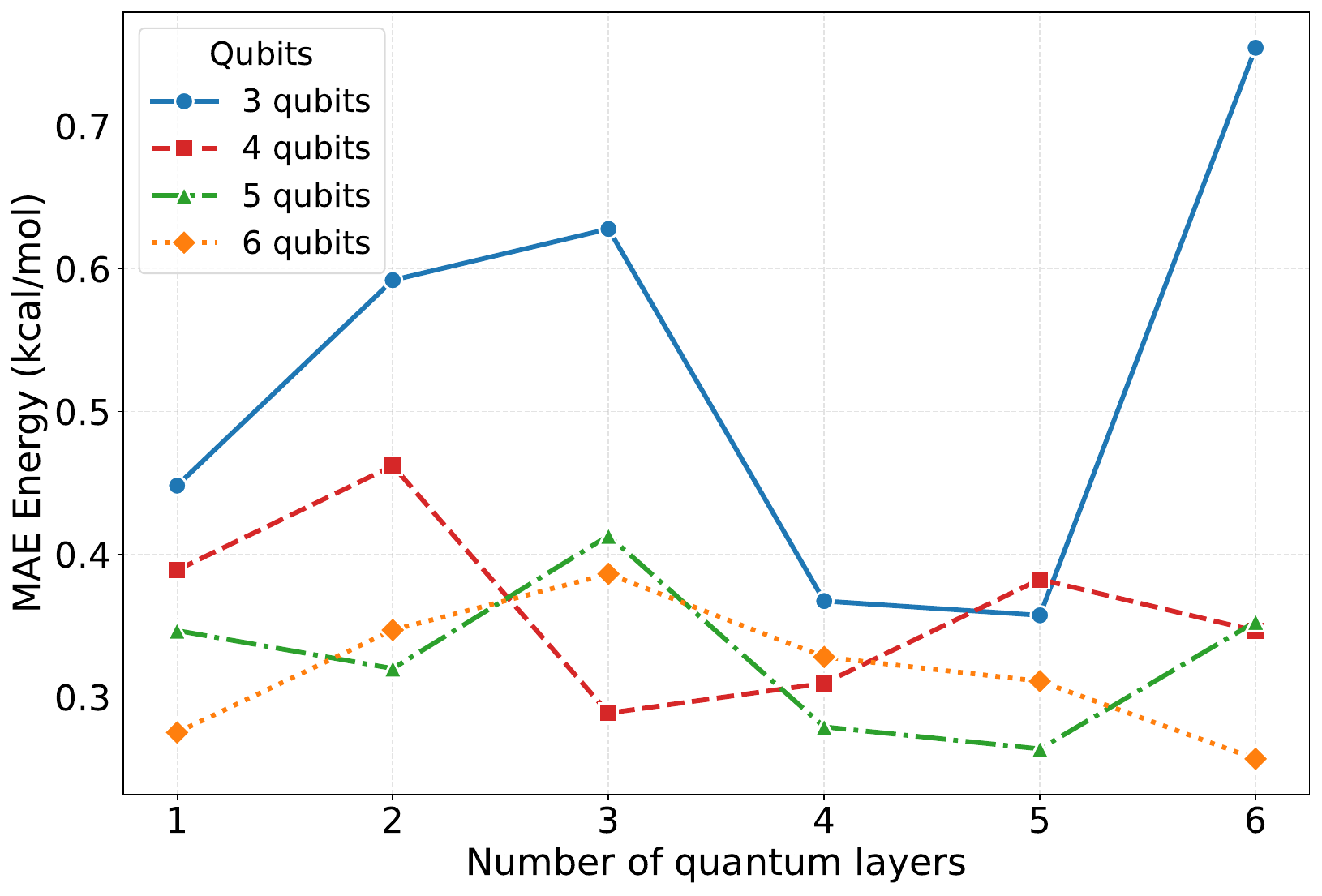}
        \caption{MAE energy experiments of ethanol across individual layer configurations.}
        \label{fig:ethanol_mae_layers}
    \end{subfigure}
    \caption{MAE energy experiments of ethanol for different quantum circuit configurations.}
    \label{fig:ethanol_mae_experiments}
\end{figure}

The proposed Hybrid Quantum SchNet model is evaluated on the MD17 benchmark across eight molecular systems: benzene, toluene, malonaldehyde, salicylic acid, aspirin, ethanol, uracil, and naphthalene. Each molecule is trained independently with $N=1000$ training configurations. The model receives atomic numbers and Cartesian coordinates as input, predicts a scalar molecular energy, and computes atomic forces from the negative gradient of the predicted energy with respect to atomic positions. This construction preserves an energy-conserving force field while allowing the classical SchNet backbone and the variational quantum modules to be optimized jointly.

Two supervision settings are compared. In the energy-only setting, the loss uses $\lambda_E=1.0$ and $\lambda_F=0.0$, so the model is optimized only against molecular energies. In the joint energy--force setting, the loss uses $\lambda_E=0.01$ and $\lambda_F=1.0$, placing the dominant training signal on atomic forces while retaining an energy constraint. All models are trained with Adam~\cite{kingma2017adammethodstochasticoptimization}, a batch size of 32, a decay rate of 0.96, and an exponential moving average update after each optimization step. Separate learning rates are used for the two parameter groups: $10^{-3}$ for the classical parameters and $10^{-4}$ for the quantum circuit parameters.

Figure~\ref{fig:ethanol_mae_experiments} reports the ethanol ablation study over quantum circuit width and depth. The trend shows that predictive accuracy is not controlled by a single circuit hyperparameter. Instead, the strongest configurations reflect a balance between the expressive capacity of the quantum transformation and the stability of the optimization process. Increasing the number of qubits or strongly entangling layers can enlarge the representational capacity of the hybrid module, but deeper circuits do not automatically reduce MAE. This behavior is consistent with the known optimization difficulty of variational quantum circuits and suggests that quantum circuit design for force fields must consider trainability as carefully as expressivity.

\subsection{\label{sec:prediction_metrics}Energy and force prediction performance}

Table~\ref{tab:hybrid_SchNet_md17_results} compares the proposed Hybrid Quantum SchNet with our reproduced classical SchNet baseline on the MD17 benchmark using $N=1000$ training configurations. Both models were trained and evaluated using the same experimental protocol and data partition, allowing a direct comparison between the classical and hybrid parameterizations.

Under energy-only training, the hybrid model gives comparable or lower force errors for several molecules. For benzene, the force MAE decreases from 14.12 to 6.104 kcal mol$^{-1}$ \AA$^{-1}$, while for aspirin it decreases from 23.54 to 19.840 kcal mol$^{-1}$ \AA$^{-1}$. These values should not be interpreted as evidence that the hybrid model is generally superior to classical SchNet. In the energy-only setting, forces are obtained only as gradients of an energy model and are not directly supervised; moreover, the classical values are taken from the reference implementation in the original paper. The energy-only comparison is therefore mainly diagnostic, showing that the hybrid model preserves a differentiable energy--force relation, while the molecule-dependent behavior remains substantial.

When force labels are included, the classical SchNet baseline is consistently stronger across the MD17 molecules in this comparison. The hybrid model still benefits substantially from joint energy--force training, but its MAE values remain higher than those of the fully classical SchNet reported in the original benchmark. This is the key message of Table~\ref{tab:hybrid_SchNet_md17_results}: the proposed quantum-enhanced model is trainable and physically consistent, but it does not yet outperform the mature classical SchNet architecture. The objective of the present work is therefore not to claim state-of-the-art accuracy, but to evaluate whether variational quantum circuit modules can be embedded into a constrained neural force field and optimized end-to-end.

The comparison therefore provides a more informative baseline than reporting the hybrid model alone. Classical SchNet establishes the expected accuracy of the continuous-filter architecture on MD17, whereas Hybrid Quantum SchNet tests whether selected nonlinear transformations in that architecture can be replaced by VQC-based modules while preserving trainability. The remaining gap under joint force supervision suggests that future work should investigate circuit ansatz design, quantum-module placement, initialization, and optimization schedules before the hybrid model can be considered competitive with optimized classical neural potentials.

\subsection{\label{sec:training_dynamics}Training dynamics}

\begin{figure}[!htbp]
    \centering
    \includegraphics[width=0.85\linewidth]{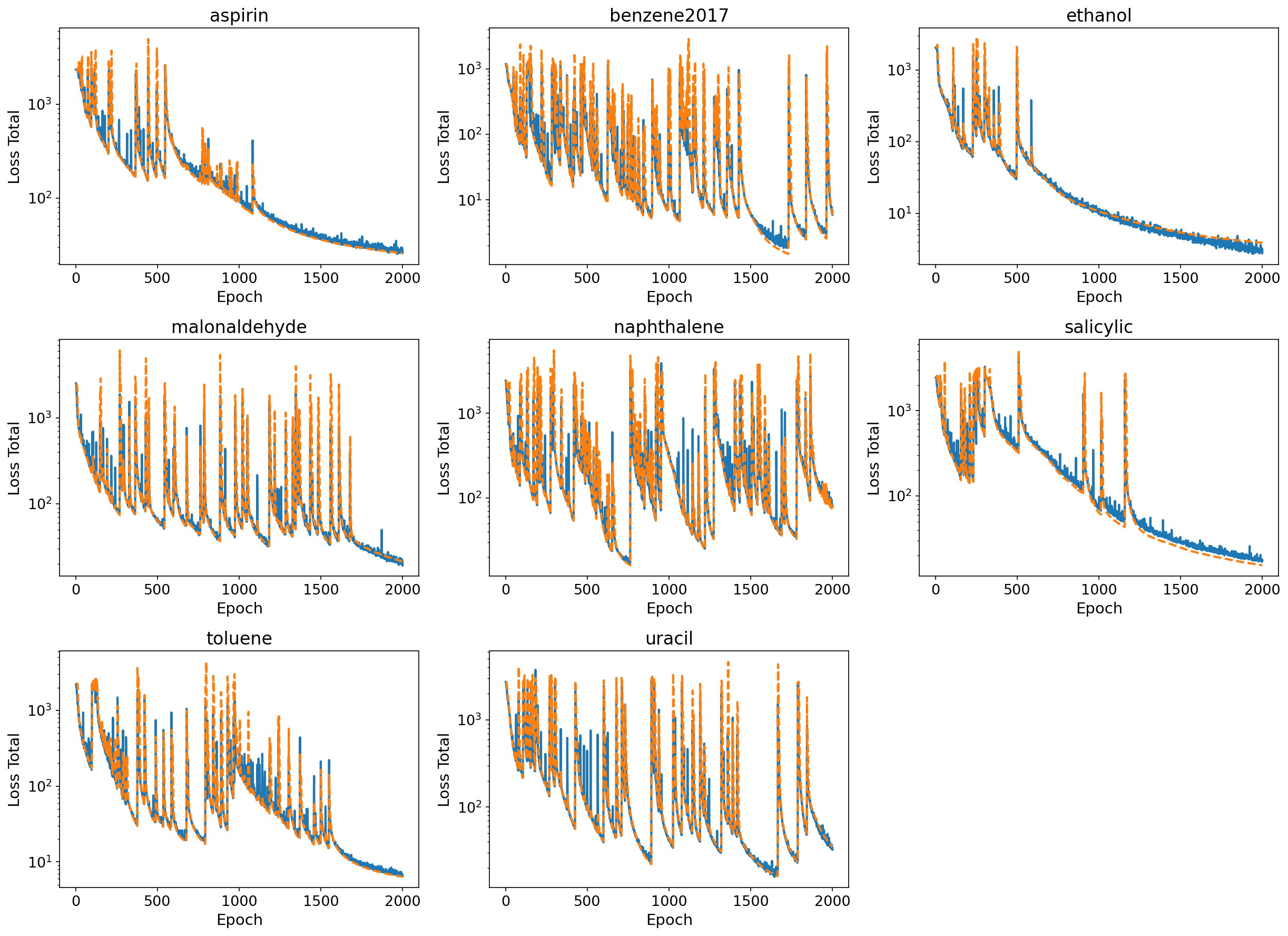}
    \caption{
Training and validation total loss curves for aspirin, benzene, ethanol, and malonaldehyde on the MD17 benchmark. 
The total loss is plotted on a logarithmic scale as a function of training epoch. 
The blue solid curves denote the training loss, while the orange dashed curves denote the validation loss.
}
\label{fig:loss_curves_log_md17_part1}
\end{figure}

Figure~\ref{fig:loss_curves_log_md17_part1} shows the training and validation total loss curves for aspirin, benzene, ethanol, and malonaldehyde. The losses are plotted on a logarithmic scale, making it possible to compare the rapid early-stage decrease with the slower refinement observed at later epochs. Across the four molecules, the validation curves generally follow the same downward trend as the training curves, indicating that the hybrid model does not simply memorize the training configurations but learns patterns that transfer to unseen molecular geometries.

The convergence behavior remains stable despite the inclusion of variational quantum circuit modules. Ethanol displays the smoothest optimization trajectory, with both training and validation losses decreasing steadily. Aspirin and malonaldehyde also show clear long-term improvement, although their curves contain stronger early fluctuations. Benzene exhibits sharper oscillations, but the validation loss remains coupled to the training loss rather than diverging from it. This suggests that the oscillations are mainly associated with optimization dynamics rather than severe overfitting. From a hybrid quantum-classical perspective, this behavior is important because VQC parameters can introduce additional nonconvexity and gradient variability. The observed convergence indicates that the shallow circuit design, classical pre- and post-processing layers, and separate learning rates provide a practical training regime for the proposed architecture.

\subsection{\label{sec:discussion_implications}Discussion and implications}

Taken together, the numerical results support three main observations. First, force supervision is essential for learning an energy-conserving potential that is useful beyond scalar energy regression. The largest improvements occur in the force metrics, confirming that the local geometry of the learned potential energy surface is strongly shaped by direct gradient supervision. Second, the proposed architecture demonstrates that VQC modules can be inserted into a SchNet-like force field without destroying end-to-end trainability. This is not a trivial outcome, because the hybrid model must propagate gradients through classical interaction blocks, quantum expectation values, and energy-gradient force calculations. Third, the ethanol ablation results suggest that quantum circuit design should be treated as a coupled capacity--optimization trade-off. Larger circuits may offer richer transformations, but the realized benefit depends on whether the training procedure can exploit the additional parameters.

These findings should be interpreted as a feasibility study rather than a final performance benchmark against mature equivariant neural potentials. The main scientific value of the present results is that they identify a physically consistent route for incorporating variational quantum circuits into neural force-field architectures. The SchNet backbone supplies the necessary molecular inductive biases, while the quantum modules provide alternative trainable nonlinear maps inside the filter, interaction, and readout components. This separation of roles makes the framework suitable for future studies of quantum-module placement, circuit ansatz design, measurement strategies, and hybrid optimization protocols.

\section{\label{sec:conclusions}Conclusions}

In this work, we presented Hybrid Quantum SchNet, a hybrid quantum-classical neural force-field architecture for molecular energy and force prediction. The model preserves the physically motivated structure of SchNet while introducing variational quantum circuit modules into selected trainable components, including the continuous-filter generator, the atom-wise interaction transformation, and the final readout network. By representing the molecular energy as a sum of atom-wise contributions and computing atomic forces as negative energy gradients, the proposed model retains an energy-conserving formulation while allowing quantum-enhanced nonlinear transformations to participate in the learning process.

Experiments on the MD17 benchmark show that the hybrid architecture can be trained end-to-end using molecular energy and force data. Across eight molecules with $N=1000$ training configurations, joint energy--force supervision consistently improves over energy-only training. On average, the energy MAE decreases from 2.567 to 0.593 kcal mol$^{-1}$, while the force MAE decreases from 16.340 to 1.540 kcal mol$^{-1}$ \AA$^{-1}$. These results confirm that force supervision provides essential local-gradient information for learning accurate potential energy surfaces and reliable molecular force fields.

The ethanol ablation study further shows that hybrid quantum circuit design involves a trade-off between expressivity and trainability. Increasing the number of qubits or strongly entangling layers can improve the capacity of the VQC module, but larger circuits do not automatically yield lower prediction error. This observation is consistent with the optimization challenges of variational quantum circuits and suggests that circuit architecture, initialization, measurement design, and learning-rate schedules should be considered jointly when developing quantum-enhanced force fields.

Overall, the results support the feasibility of embedding variational quantum circuits within a symmetry-aware neural force-field architecture. The proposed framework is not intended to replace state-of-the-art equivariant classical potentials at this stage. Instead, it provides a controlled and physically consistent platform for studying how quantum modules behave when they are coupled to distance-based molecular interactions and energy-gradient force prediction. Future work will investigate larger molecular systems, more diverse chemical environments, systematic ablations of individual quantum-module placements, more expressive yet trainable ansatz families, and strategies for reducing the computational overhead of hybrid quantum-classical force-field training.

\bibliography{Main}


\end{document}